\documentclass[aps,twocolumn,a4paper,floatfix,tightenlines,superscriptaddress,pra,longbibliography]{revtex4-1}
\usepackage[toc,page]{appendix}
\usepackage{braket}
\usepackage{epsfig,graphicx,times}
\usepackage{amstext}
\usepackage{amsmath}
\usepackage{amssymb}
\usepackage{mathrsfs}
\usepackage{dcolumn}
\usepackage{bm}
\usepackage[tight]{subfigure}
\usepackage[colorlinks,linkcolor=red,anchorcolor=blue,citecolor=blue,urlcolor=red]{hyperref}
\usepackage{color}
\usepackage{siunitx}
\usepackage{appendix}
\usepackage{placeins}
\usepackage{textcomp}
\usepackage{bbold}
\usepackage{float}
\usepackage{verbatim}
\usepackage{minitoc}
\usepackage[dvipsnames]{xcolor}

\usepackage{soul}
\usepackage[framemethod=tikz]{mdframed}
\usepackage{CJKutf8}

\begin{document}
\begin{CJK}{UTF8}{<font>}

\title{Reversible optical-microwave quantum conversion assisted by \\ optomechanical dynamically-dark modes}

\author{Ling-Ying Zhu}
\affiliation{Key Laboratory of Low-Dimensional Quantum Structures and Quantum Control of Ministry of Education, Department of Physics and Synergetic Innovation Center for Quantum Effects and Applications, Hunan Normal University, Changsha 410081, China}

\author{Yong Dong}
\affiliation{Key Laboratory of Low-Dimensional Quantum Structures and Quantum Control of Ministry of Education, Department of Physics and Synergetic Innovation Center for Quantum Effects and Applications, Hunan Normal University, Changsha 410081, China}

\author{Ji Zhang}
\affiliation{Key Laboratory of Low-Dimensional Quantum Structures and Quantum Control of Ministry of Education, Department of Physics and Synergetic Innovation Center for Quantum Effects and Applications, Hunan Normal University, Changsha 410081, China}

\author{ Cui-Lu Zhai}
\affiliation{Key Laboratory of Low-Dimensional Quantum Structures and Quantum Control of Ministry of Education, Department of Physics and Synergetic Innovation Center for Quantum Effects and Applications, Hunan Normal University, Changsha 410081, China}

\author{Le-Man Kuang}\email{lmkuang@hunnu.edu.cn}
\affiliation{Key Laboratory of Low-Dimensional Quantum Structures and Quantum Control of Ministry of Education, Department of Physics and Synergetic Innovation Center for Quantum Effects and Applications, Hunan Normal University, Changsha 410081, China}

\begin{abstract}

We propose a dynamically-dark-mode (DDM)  scheme to realize the reversible quantum conversion between microwave and optical photons in an electro-optomechanical (EOM) model. It is shown that two DDMs appear at certain times during the dynamical evolution of the EOM model.  It is demonstrated that the DDMs can induce two kinds of  reversible and highly efficient quantum conversion between the microwave and optical fields, the conditional quantum conversion (CQC) and the entanglement-assisted quantum conversion (EAQC). The CQC happens at the condition of vanishing of the initial-state mean value of one of the microwave and optical fields, and only depends on the coupling ratio of the system under consideration. The EAQC occurs in the presence of the initial-state entanglement  between the microwave and optical fields. It is found  that  the EAQC can be manipulated by engineering the initial-state entanglement and the coupling ratio. It is indicated that it is possible to realize the entanglement-enhanced (or suppressed) quantum conversion through controlling the phase of the initial-state parameter. Our work highlights the power of generating reversible and highly efficient quantum conversion between microwave and optical photons by the DDMs.

\end{abstract}

\maketitle

\section{Introduction}

In recent years, much attention has been paid to quantum conversion between microwave and optical photons due to its importance in quantum technologies \cite{Zeuthen,Lambert,Lauk}. Such quantum conversion is of special significance to realize a quantum internet \cite{Kimble,Castelvecchi,Reiserer,Wehner} and distributed quantum tasks including computing or sensing \cite{Pirandola,Maccone}.
A number of systems such as atomic, molecular, and solid-state impurity spins \cite{Sorensen,TianRabl,RablDeMille,OBrien,XiaTwamley,Das,Gard,Lekavicius}, magnons in ferromagnetic materials \cite{Hisatomi}, electro-optic modulators \cite{Tsang1,Tsang2,JaverzacGaly,Rueda}, and mechanical oscillators \cite{Stannigel,Taylor,Barzanjeh,Tian,WangClerk,Clader,YinYang,Hammerer,Okada}, have been proposed as suitable candidates for mediating interaction between microwave and optical fields. In particular,  the electro-optomechanical (EOM) system based on quantum cavity optomechanics  is regarded  as a promising and versatile platform with several experiments demonstrating efficient conversion between the microwave and optical fields \cite{Bochmann,Andrews,Bagci,Balram}.

The mechanical modes in cavity optomechanical systems \cite{Bowen,Aspelmeyer,Kippenberg,Jiao,Tan,Zhai} can couple with the optical and microwave modes. The   coupling between the mechanical mode and the optical and microwave modes has been  demonstrated in recent experiments \cite{Groblacher,Weis,Safavi-Naeini,Chan,Brahms,Verhagen,Massel,Agarwal,Zhou,OConnell,Teufel,Riviere,Thompson}. Such systems can hence serve as an interface in quantum networks to connect optical and microwave photons \cite{Cirac}.
It is well known that the optomechanical dark mode exists in a lot of cavity optomechanical systems \cite{Dong,Wang,Tian1,Zhang}. The optomechanical dark mode was experimentally demonstrated by coupling two optical whispering gallery modes to a mechanical breathing mode in a silica resonator in the regime of weak optomechanical coupling  \cite{Dong}.
Since the optomechanical dark mode can protect the system from mechanical dissipation, it can be employed for the realization of high efficient quantum state conversion \cite{Wang,Tian1}, the reservoir-engineered entanglement \cite{Wang2,Tian2}, quantum illumination \cite{Barzanjeh},  optomechanically induced transparency  \cite{Liao,Lake}, and  wider quantum applications \cite{Kuzyk,Sommer,Ockeloen-Korppi}.

In this work we theoretically investigate the reversible optical-to-microwave quantum conversion arising from the optomechanical dynamically-dark modes (DDMs) which are dynamically decoupled from the mechanical resonator at some specific times in an EOM quantum conversion model. The presence of the DDMs results in a bidirectional and highly efficient quantum conversion  between microwave and optical fields.
The remainder of the paper is organized as follows. In Sec. II, we introduce the electro-optomechanical (EOM) quantum conversion model,  present an approximately analytical solution to the EOM model, and demonstrate the existence of the dynamically-dark mode.  In Sec. III, we show reversible and highly efficient quantum conversion between the optical and microwave fields.  In Sec. IV, we study the entanglement-assisted quantum conversion between the optical and microwave fields. We show that the DDMs can induce two kinds of  bidirectional and highly efficient quantum conversion between the microwave and optical fields, the conditional quantum conversion  and the entanglement-assisted quantum conversion. Finally, the concluding section, Sec. V, summarizes and  discusses our main results.

\section{The electro-optomechanical model and dynamically-dark modes}

In this section we introduce the cavity EOM model and present an approximately analytical solution of this model in terms of the Senm-Mandal method \cite{sm1,sm2,sm3}.
The  cavity EOM model under our consideration is a EOM quantum converter which consists of a driven superconducting microwave cavity of resonant frequency $\omega_w$, a driven a Fabry-P\'{e}rot optical cavity  with resonant frequency $\omega_o$, and a mechanical resonator with resonant frequency $\omega_m$ \cite{BarzanjehGuha,BarzanjehVitali}.
The mechanical resonator  (annihilation operator $\hat{b}$) is capacitively coupled on one side to a driven superconducting microwave cavity (annihilation operator $\hat{a}_w$), and on the other side to a driven Fabry-Perot optical cavity (annihilation operator $\hat{a}_o$)  \cite{Bochmann,Andrews,Bagci,BarzanjehGuha,BarzanjehVitali}. In the frame rotating at the frequencies of the microwave and optical driving fields, the Hamiltonian of the EOM model \cite{BarzanjehGuha}  is given by
\begin{eqnarray}
\hat{H}&=&\hbar\omega_m \hat{b}^{\dag}\hat{b} + \hbar\sum_{j=w,o}\left[\Delta_{j}+g_j(\hat{b} +\hat{b}^{\dag})\right]\hat{a}^{\dag}_j\hat{a}_j +\hbar\sum_{j=w,o}E_j(\hat{a}^{\dag}_j-\hat{a}_j ),
\end{eqnarray}
where $g_j$ is the coupling constant between the mechanical resonator and cavity $j$, $\Delta_{j}=\omega_j-\omega_{d,j}$ are  the detunings from their resonant frequencies $\omega_j$ with $j=w,o$ denoting the microwave and optical cavities,   $ E_{o}$ and $ E_{w}$ are the optical and microwave driving field amplitudes \cite{BarzanjehVitali}, respectively.

we can treat cavity modes with semi-classical description and linearize the Hamiltonian by expanding the cavity modes around their steady-state field amplitudes $\hat{c}_j=\hat{a}_j-\sqrt{N_j}$ with the $N_j=|E_j|^2/(\kappa^2_j+\Delta^2_j) \gg 1$ being the mean numbers of intracavity photons induced by the microwave or optical pumps \cite{BarzanjehVitali}  with   $\kappa_j$ are the total cavity decay rates, since the single photon coupling constants between the optical and microwave  fields  and the mechanical resonator  $g_j$  are small in current experiments \cite{Barzanjeh}. When we choose the effective cavity detunings $\Delta_{w}=-\Delta_{o}=\omega_m$ and neglect the terms at $\pm\omega_m$, under rotating wave approximation the linearized Hamiltonian of the EOM model is given by
\begin{eqnarray}
\hat{H}=\hbar G_o(\hat{c}_o\hat{b} +\hat{b}^{\dag}\hat{c}^{\dag}_o) +\hbar G_w(\hat{c}_w\hat{b}^{\dag} +\hat{b}\hat{c}^{\dag}_w),
\end{eqnarray}
where the multiphoton coupling rate is $G_j=g_j\sqrt{N_j}$. We have described above interaction Hamiltonian in Fig. 1.

\begin{figure}[htp]
\centering
\includegraphics[width=7.0cm,height=6.2cm]{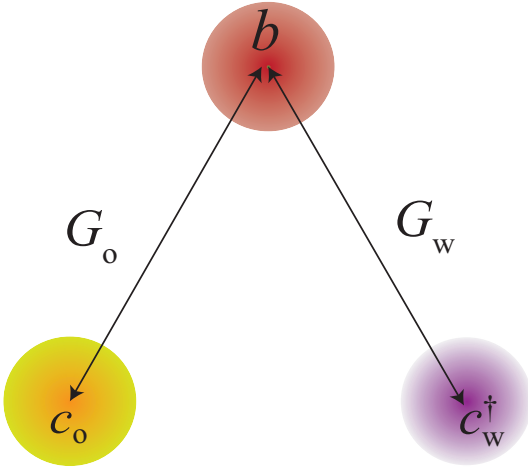}
\caption{(Color online) The schematic diagram of the electro-optomechanical quantum conversion model. }\label{fig1}
\end{figure}

In what follows we solve the above linearized Hamiltonian in the Heisenberg picture by means of the Senm-Mandal method \cite{sm1,sm2,sm3}. From the Hamiltonian (2) we can obtain the Heisenberg operator equations of motion involving the field operators
\begin{eqnarray}
\dot{\hat{b}}(t)&=& -iG_{o}\hat{c}^{\dagger}_{o}(t)-iG_{w}\hat{c}_{w}(t),\nonumber \\
\dot{\hat{c}}_{o}(t)&=& -iG_{o}\hat{b}^{\dagger}(t), \nonumber \\
\dot{\hat{c}}_{w}(t)&=&-iG_{w}\hat{b}(t).
\end{eqnarray}

Note that the field operators in the Heisenberg picture have the following formal solution
\begin{eqnarray}
\hat{b}(t)&=&\exp(i\hat{H}t)\hat{b}(0)\exp(-i\hat{H}t), \nonumber \\
\hat{c}_o(t)&=&\exp(i\hat{H}t)\hat{c}_o(0)\exp(-i\hat{H}t), \nonumber \\
\hat{c}_w(t)&=&\exp(i\hat{H}t)\hat{c}_w(0)\exp(-i\hat{H}t),
\end{eqnarray}
which can be expanded as
\begin{eqnarray}
\hat{b}(t) &=&\hat{b}(0)+it\left[\hat{H},\hat{b}(0)\right]+(it)^{2}\left[\hat{H},\left[\hat{H},\hat{b}(0)\right]\right]\nonumber \\
& &+(it)^{3}\left[\hat{H},\left[\hat{H}, \left[\hat{H},\hat{b}(0)\right]\right]\right]+ \cdots \nonumber \\
\hat{c}_o(t) &=&\hat{c}_o(0)+it\left[\hat{H},\hat{c}_o(0)\right]+(it)^{2}\left[\hat{H},\left[\hat{H},\hat{c}_o(0)\right]\right]\nonumber \\
& &+(it)^{3}\left[\hat{H},\left[\hat{H}, \left[\hat{H},\hat{c}_o(0)\right]\right]\right]+ \cdots \nonumber \\
\hat{c}_w(t) &=&\hat{c}_w(0)+it\left[\hat{H},\hat{c}_w(0)\right]+(it)^{2}\left[\hat{H}, \left[\hat{H},\hat{c}_w(0)\right]\right]\nonumber \\
& &+(it)^{3}\left[\hat{H}, \left[\hat{H}, \left[\hat{H},\hat{c}_w(0)\right]\right]\right]+ \cdots
\end{eqnarray}

For the mechanical mode, the related   commutators are obtained as follows
\begin{eqnarray}
\left[\hat{H}, \hat{b}(0)\right]&=&-G_{o}\hat{c}_{o}^{\dagger}(0)-G_{w}\hat{c}_{w}(0), \nonumber \\
\left[\hat{H}, \left[\hat{H}, \hat{b}(0)\right]\right]&=&-G^{2}_{o}\hat{b}(0)+G^{2}_{w}\hat{b}(0), \nonumber \\
\left[\hat{H},\left[\hat{H},\left[\hat{H},\hat{b}(0)\right]\right]\right] &=&(G^{2}_{o}-G^{2}_{w})[G_{o}\hat{c}_{o}^{\dagger}(0)+G_{w}\hat{c}_{w}(0)]. \nonumber \\
\end{eqnarray}

It is straightforward to show that all higher-order commutators in the right hand side of the first equation in Eq. (5) contain only the three initial operators $\hat{b}(0)$, $\hat{c}_{w}(0)$  and  $\hat{c}_{o}^{\dagger}(0)$.
Then the operator of the mechanical mode $\hat{b}(t)$ is a linear superposition of the three operators $\hat{b}(0)$, $\hat{c}_{w}(0)$  and  $\hat{c}_{o}^{\dagger}(0)$. Similarly, we find that the operator  of the optical mode $\hat{c}_o(t)$ is a linear superposition of the three initial operators $\hat{c}_o(0)$, $\hat{c}^{\dag}_{w}(0)$  and  $\hat{b}^{\dagger}(0)$ while the operator  of the microwave mode $\hat{c}_w(t)$ is a linear superposition of the three initial operators $\hat{c}_w(0)$, $\hat{c}^{\dag}_{o}(0)$  and  $\hat{b}(0)$.
Therefore,  the approximate analytical solution of the Heisenberg equation of motion can be directly expressed as
\begin{eqnarray}
\hat{b}(t)    &=& f_{1}(t)\hat{b}(0)+f_{2}(t)\hat{c}_{w}(0)+f_{3}(t)\hat{c}_{o}^{\dagger}(0),\nonumber \\
\hat{c}_{o}(t)&=& g_{1}(t)\hat{c}_{o}(0)+g_{2}(t)\hat{c}_{w}^{\dagger}(0)+g_{3}(t)\hat{b}^{\dagger}(0),  \nonumber \\
\hat{c}_{w}(t)&=& h_{1}(t)\hat{c}_{w}(0)+h_{2}(t)\hat{c}_{o}^{\dagger}(0)+h_{3}(t)\hat{b}(0),
\label{evolution 1}
\end{eqnarray}
where  the time-dependent coefficient functions $f_{i}(t)$, $g_{i}(t)$, $h_{i}(t)$, $(i=1,2,3)$ should satisfy the following initial conditions
\begin{eqnarray}
f_{1}(0)= g_{1}(0)=h_{1}(0)&=&1,  \nonumber \\
f_{2}(0)= g_{2}(0)=h_{2}(0)&=&0,  \nonumber \\
f_{3}(0)= g_{3}(0)=h_{3}(0)&=&0.
\label{evolution 1}
\end{eqnarray}

The self-consistency of the solutions leads to
\begin{eqnarray}
\left[\hat{b}(t),\hat{b}^{\dagger}(t)\right]&=&|f_{1}(t)|^{2}+|f_{2}(t)|^{2}-|f_{3}(t)|^{2}=1, \nonumber \\
\left[\hat{c}_{o}(t),\hat{c}_{o}^{\dagger}(t)\right]&=&  |g_{1}(t)|^{2}-|g_{2}(t)|^{2}-|g_{3}(t)|^{2}=1, \nonumber \\
\left[\hat{c}_{w}(t),\hat{c}_{w}^{\dagger}(t)\right]&=& |h_{1}(t)|^{2}-|h_{2}(t)|^{2}+|h_{3}(t)|^{2}=1,
\end{eqnarray}
which indicates that at any time all of the three isochronal commutation relations satisfy  commutation relations of  the harmonic oscillator to ensure the self-consistency of the three mode solutions.

Substituting Eq. (7) into Eq. (3) and  comparing the coefficients of equations, we obtain the differential equations of the coefficient functions
\begin{eqnarray}
\dot{f}_{1}(t)&=& -iG_{o}g_{3}^{\ast}(t)-iG_{w}h_{3}(t),\nonumber \\
\dot{f}_{2}(t)&=& -iG_{o}g_{2}^{\ast}(t)-iG_{w}h_{1}(t),\nonumber \\
\dot{f}_{3}(t)&=& -iG_{o}g_{1}^{\ast}(t)-iG_{w}h_{2}(t),\nonumber \\
\dot{g}_{1}(t)&=& -iG_{o}f_{3}^{\ast}(t), \hspace{0.5cm}
\dot{g}_{2}(t)= -iG_{o}f_{2}^{\ast}(t), \nonumber \\
\dot{g}_{3}(t)&=& -iG_{o}f_{1}^{\ast}(t),\nonumber \\
\dot{h}_{1}(t)&=& -iG_{w}f_{2}(t),\hspace{0.5cm}
\dot{h}_{2}(t) = -iG_{w}f_{3}(t),\nonumber \\
\dot{h}_{3}(t)&=& -iG_{w}f_{1}(t).
\end{eqnarray}

For the above  differential equations, when $G_{w}\neq G_{o}$, we can obtain the following  solution
\begin{eqnarray}
f_{1}(t)&=& \cos\Omega t, \hspace{0.5cm}
f_{2}(t)  = -i\frac{\sin\Omega t}{\sqrt{1-k^{2}}}, \hspace{0.5cm}
f_{3}(t)= kf_{2}(t), \nonumber \\
g_{1}(t)&=& -\frac{k^{2}\cos\Omega t}{1-k^{2}}+\frac{1}{1-k^{2}},\nonumber \\
g_{2}(t)&=& -\frac{k\cos\Omega t}{1-k^{2}}+\frac{k}{1-k^{2}}, \nonumber \\
g_{3}(t)&=& -i\frac{k\sin\Omega t}{\sqrt{1-k^{2}}}, \nonumber \\
h_{1}(t)&=& \frac{\cos\Omega t}{1-k^{2}}-\frac{k^{2}}{1-k^{2}},\hspace{0.5cm}
h_{2}(t)= \frac{k\cos\Omega t}{1-k^{2}} -\frac{k}{1-k^{2}} ,\nonumber\\
h_{3}(t)&=& -i\frac{\sin\Omega t}{\sqrt{1-k^{2}}},
\label{evolution 1}
\end{eqnarray}
where we we have introduced the ratio of two coupling strengths $k$ and $\Omega$ defined by
\begin{equation}
k=\frac{G_{o}}{G_{w}}, \hspace{0.5cm} \Omega=\sqrt{G_{w}^{2}-G_{o}^{2}}=\sqrt{1-k^{2}}G_{w}.
\end{equation}

According to the solution given by Eq. (11), it is interesting to note that the optical-wave and microwave can be decoupled with mechanical vibrator at some special times. In fact,  in equation (13) if we choose special moments
\begin{equation}
t=t_{n}=\frac{n\pi}{\Omega}, \hspace{0.5cm} n=0,1,2,3, \cdots
 \end{equation}
the solution given by Eq. (11) becomes
\begin{eqnarray}
f_{1}(t_{n})&=& -1,\hspace{0.5cm}
f_{2}(t_{n})= f_{3}(t_{n})= 0, \nonumber \\
g_{1}(t_{n})&=& \frac{1+k^{2}}{1-k^{2}},\hspace{0.5cm}
g_{2}(t_{n})= \frac{2k}{1-k^{2}},\hspace{0.5cm}
g_{3}(t_{n})= 0 \nonumber \\
h_{1}(t_{n})&=& -\frac{1+k^{2}}{1-k^{2}},\hspace{0.5cm}
h_{2}(t_{n})=-\frac{2k}{1-k^{2}}, \nonumber \\
h_{3}(t_{n})&=& 0.
\label{evolution 1}
\end{eqnarray}
Obviously, these coefficient functions  satisfy the following relationship
\begin{equation}
|g_{1}(t_{n})|^{2}-|g_{2}(t_{n})|^{2}=1, \hspace{0.5cm} |h_{1}(t_{n})|^{2}-|h_{2}(t_{n})|^{2}=1.
\end{equation}

By substituting Eq. (14) into Eq. (7),  we can rewrite the solution of the EOM model as
\begin{eqnarray}
\hat{c}_{o}(t_{n})&=& g_{1}(t_{n})\hat{c}_{o}(0)+g_{2}(t_{n})\hat{c}_{w}^{\dagger}(0),  \nonumber \\
\hat{c}_{w}(t_{n})&=& h_{1}(t_{n})\hat{c}_{w}(0)+h_{2}(t_{n})\hat{c}_{o}^{\dagger}(0),
\label{evolution 1}
\end{eqnarray}
which indicate that the optical-wave (microwave) field mode at the time $t_{n}$  only involves the initial optical-field operator  and the initial microwave-field operator.
Hence, the optical-wave and microwave modes are well decoupled with the mechanical oscillator at the moments $t=t_{n}$, in this sense we call the two modes $\hat{c}_{o}(t_{n})$
and $\hat{c}_{w}(t_{n})$ as dynamically-dark modes with respect to the mechanical mode.

These DDMs are mechanically-dark modes which are superpositions of the optical and microwave modes. Although they are decoupled from the mechanical oscillator, they can still mediate an effective optomechanical coupling between the optical and microwave modes. In the following we will investigate the DDM-assisted quantum conversion between the optical and microwave fields.

\section{Reversible and highly efficient optical-to-microwave  quantum conversion }

In this section we investigate the DDM-assisted quantum conversion between optical and microwave fields in the EOM converter. In general, a full description of the quantum conversion system should includes the inputs and outputs of the optical, microwave and mechanical resonators. However, we only need to pay our attention  to the inputs and outputs of the optical and microwave fields in the DDM case since the optical and microwave fields are completely decoupled to the  mechanical resonator. The EOM Converter performance \cite{Andrews}  can be characterized by the quantum conversion rate between the optical (microwave) and  microwave (optical) fields defined by
\begin{equation}
\eta_{ow}=\left|\frac{\langle \hat{c}_{w}(t_n)\rangle}{\langle \hat{c}_{o}(0)\rangle}\right|^{2}, \hspace{0.5cm}
\eta_{wo}=\left|\frac{\langle \hat{c}_{o}(t_n)\rangle}{\langle \hat{c}_{w}(0)\rangle}\right|^{2}.
\end{equation}

Substituting Eq. (16) into Eq. (17) we obtain
\begin{eqnarray}
\eta_{ow}&=&\left|h_1(t_n)\frac{\langle \hat{c}_{w}(0)\rangle}{\langle \hat{c}_{o}(0)\rangle} + h_2(t_n)\frac{\langle \hat{c}^{\dag}_{o}(0)\rangle}{\langle \hat{c}_{o}(0)\rangle} \right|^{2}, \nonumber\\
\eta_{wo}&=&\left|g_1(t_n)\frac{\langle \hat{c}_{o}(0)\rangle}{\langle \hat{c}_{w}(0)\rangle} + g_2(t_n)\frac{\langle \hat{c}^{\dag}_{w}(0)\rangle}{\langle \hat{c}_{w}(0)\rangle} \right|^{2},
\end{eqnarray}
which indicates that when the initial-state mean value of the output mode vanishes, i.e., $\langle \hat{c}_{w}(0)\rangle=0$ ($\langle \hat{c}_{o}(0)\rangle=0$), the quantum conversion rate from the input optical (microwave) to output microwave (optical) fields is simply reduced to
\begin{eqnarray}
\eta_{ow}&=&\left|h_2(t_n)\right|^{2}=\frac{4k^2}{(1-k^2)^2}, \nonumber\\
\eta_{wo}&=&\left|g_2(t_n)\right|^{2}=\frac{4k^2}{(1-k^2)^2}.
\end{eqnarray}

\begin{figure}[htp]
\centering
\includegraphics[width=8.0cm,height=5.0cm]{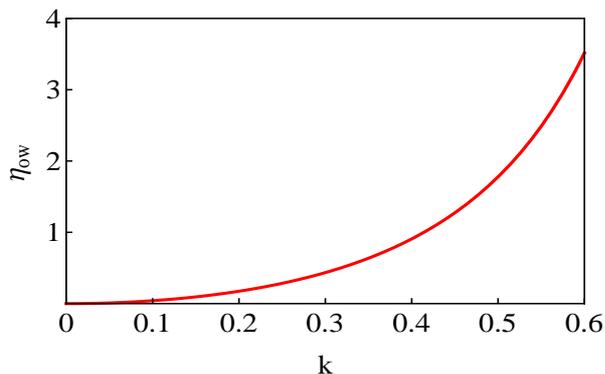}
\caption{(Color online)  The optical-to-microwave quantum conversion rate $\eta_{ow}$ with respect to the  coupling ratio $k$  for  the conditional quantum conversion between the  optical and microwave fields.} \label{fig2}
\end{figure}

It is worthwhile to note that the optical-microwave quantum conversion corresponding to Eq. (19) is a conditional quantum conversion (CQC) between the  optical and microwave fields in which the highly efficient quantum conversion from the optical (microwave) to microwave (optical) fields is at the condition of  the vanishing initial-state mean value of the microwave (optical) field. The CQC is independent of the initial state of the converted optical (microwave) field. The quantum conversion rate of the CQC depends on only the ratio of the two coupling strengths between the optical and microwave fields and the mechanical resonator under the condition of the initial-state mean value vanishing for the output mode.
From Eq. (14) we can see that the conversion rate of the CQC from the  optical field to the microwave field equals to that from the microwave field to the optical field. Therefore, we can conclude that the CQC  between the  optical and microwave fields is a reversible quantum conversion.
 In this sense the CQC can be regarded as a universal quantum conversion between the optical and microwave fields.

In order to see how to manipulate the quantum conversion rate between the optical (microwave) and  microwave (optical) fields  by varying  the   coupling ratio $k$, we have plotted the quantum conversion rate with respect to the  coupling ratio in Fig. 2. From Fig. 2 we can see that the quantum conversion rate between the optical (microwave) and  microwave (optical) fields increases with the increase of the coupling ratio.

\section{Entanglement-assisted optical-to-microwave  quantum conversion}

In this section we investigate the influence of the initial entanglement between the optical and microwave fields  on the quantum conversion rate. We will show that entanglement-assisted optical-to-microwave quantum conversion can happen in the present DDM scheme of the optical-to-microwave quantum conversion. Without loss of generality, we consider the following optical-microwave initial entangled coherent state
\begin{equation}
|\psi(0)\rangle_{ow}=N\left[\cos \theta|\alpha\rangle_o\otimes |0\rangle_w + \sin\theta|0\rangle_o\otimes |\beta\rangle_w\right],
\end{equation}
where $|\alpha\rangle_o$ and $|\beta\rangle_w$ are the Glauber coherent states, the normalization constant $N$ is given by
\begin{equation}
N^{-2}=1+\sin(2\theta)e^{-\frac{1}{2}(|\alpha|^2+|\beta|^2)}.
\end{equation}

The degree of quantum entanglement of a bipartite two-component entangled state can be described by the quantum concurrence \cite{Kuang,WangXG,Hill,Wootters}. In general, for the following bipartite two-component entangled state
\begin{equation}
|\psi\rangle=N\left[\mu|\eta\rangle\otimes |\gamma\rangle  + \nu|\xi\rangle \otimes |\delta\rangle_w\right].
\end{equation}
where $N$ is the normalization constant, the quantum concurrence is given by
\begin{equation}
C=2|\mu||\nu|N^2 \sqrt{(1-|p_1|^2)(1-|p_2|^2)},
\end{equation}
where the two state-overlapping functions $p_1$ and $p_2$ are defined by
\begin{equation}
p_1= \langle\mu|\gamma\rangle, \hspace{0.5cm} p_2= \langle\xi|\delta\rangle.
\end{equation}

Making use of the Eq. (23), we can obtain the entanglement amount of the initial entangled coherent state (20) with the following expression
\begin{equation}
C= \frac{|\sin (2\theta)|\sqrt{(1-e^{-|\alpha|^2})(1-e^{-|\beta|^2})}}{\left[1+\sin(2\theta)e^{-(|\alpha|^2+|\beta|^2)/2}\right]},
\end{equation}
which indicates that for given values of $\alpha$ and $\beta$,  when $\theta=0$ or $\pi/2$, the entanglement amount of the initial entangled coherent state (20) vanishes while when $\theta=-\pi/4$ the initial entangled coherent state (20) has the largest entanglement amount
\begin{equation}
C_{max}= \frac{\sqrt{(1-e^{-|\alpha|^2})(1-e^{-|\beta|^2})}}{\left[1-e^{-(|\alpha|^2+|\beta|^2)/2}\right]},
\end{equation}
which means that the initial entangled coherent state (20) is a maximally entangled state with $C_{max}= 1$ when $\alpha=\pm\beta$.

For the initial entangled coherent state (20), we can  obtain the mean values of the optical and microwave field operators
\begin{eqnarray}
\langle\hat{c}_{o}(0)\rangle&=& \alpha N^2\left[\cos^2\theta +\frac{1}{2}\sin(2\theta)e^{-(|\alpha|^2+|\beta|^2)/2}\right],  \nonumber\\
\langle\hat{c}_{w}(0)\rangle&=& \beta N^2\left[\sin^2\theta +\frac{1}{2}\sin(2\theta)e^{-(|\alpha|^2+|\beta|^2)/2}\right].
\end{eqnarray}

For the simplicity, we consider the case of $\alpha=\beta$. Substituting Eq. (27) into (18), we can obtain the quantum conversion rate between  the optical and microwave fields
\begin{eqnarray}
\eta_{ow}(\theta, \varphi)&=&\left|h_1(t_n)\frac{\left[2\sin^2\theta +\sin(2\theta)e^{-|\alpha|^2}\right]}{\left[2\cos^2\theta +\sin(2\theta)e^{-|\alpha|^2}\right]}\right. \nonumber\\
 &&\left. + h_2(t_n) e^{-2i\varphi} \right|^{2}, \nonumber\\
\eta_{wo}(\theta, \varphi)&=&\left|g_1(t_n)\frac{\left[2\cos^2\theta +\sin(2\theta)e^{-|\alpha|^2}\right]}{\left[2\sin^2\theta +\sin(2\theta)e^{-|\alpha|^2}\right]} \right. \nonumber\\
 &&\left. + g_2(t_n) e^{-2i\varphi} \right|^{2},
\end{eqnarray}
where we have set $\alpha=|\alpha|e^{i\varphi}$, and $h_1(t_n), h_2(t_n), g_1(t_n)$ and $g_2(t_n)$ have been given by Eq.(14).

From Eqs. (14) and (28) we can find that
\begin{eqnarray}
\eta_{ow}(\theta, \varphi)&=&\left|\frac{\left[2\sin^2\theta +\sin(2\theta)e^{-|\alpha|^2}\right]}{\left[2\cos^2\theta +\sin(2\theta)e^{-|\alpha|^2}\right]} + \frac{2k^2e^{-2i\varphi} }{1+k^2} \right|^{2}\nonumber\\
&&\times\left(\frac{1+k^2}{1-k^2}\right)^2, \nonumber\\
\eta_{wo}(\theta, \varphi) &=&\left|\frac{\left[2\cos^2\theta +\sin(2\theta)e^{-|\alpha|^2}\right]}{\left[2\sin^2\theta +\sin(2\theta)e^{-|\alpha|^2}\right]} + \frac{2k^2e^{-2i\varphi} }{1+k^2} \right|^{2}\nonumber\\
&&\times\left(\frac{1+k^2}{1-k^2}\right)^2,
\end{eqnarray}
which indicate that the optical-to-microwave quantum conversion rate depends on the initial-state parameters ($|\alpha|, \varphi$) and  the coupling ratio $k$.
In Fig.3, we have plotted the optical-to-microwave quantum conversion rate with respect to the phase of the initial-state parameter when the coupling ratio $k=0.1, 0.2, 0.6$, and $0.9$, respectively. From Fig. 3 we can see that the optical-to-microwave quantum conversion rate sensitively depends on the phase of the initial-state parameter and the coupling ratios. Hence we can conclude that the optical-to-microwave quantum conversion rate can be efficiently manipulated by varying the phase of the initial-state parameter and the coupling ratios.

\begin{figure}[htp]
\centering
\includegraphics[width=8.0cm,height=5.5cm]{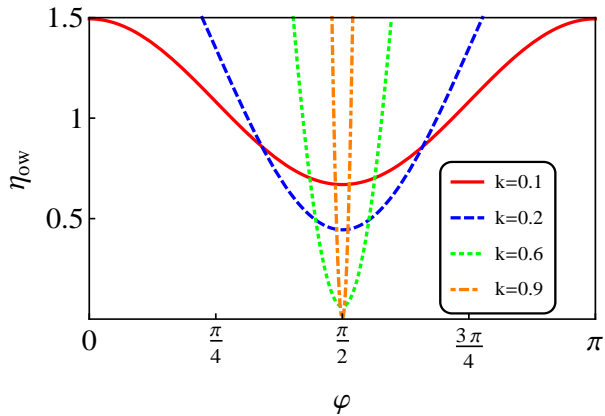}
\caption{(Color online)  The optical-to-microwave quantum conversion rate with respect to the phase of the initial-state parameter for different coupling ratios $k=0.1, 0.2, 0.6$, and $0.9$. } \label{fig3}
\end{figure}

From Eq. (29) we can find that without initial entanglement, i.e., $\cos \theta=0$  or $\sin \theta=0$, the previous quantum conversion rates given by  Eq. (19) are recovered. When the initial state (20) is an maximally entangled state with $\theta=\pi/4$, from Eq. (29) we can  find that
\begin{eqnarray}
\eta_{ow}(\pi/4, \varphi)&=&\eta_{wo}(\pi/4, \varphi)  \nonumber\\
&=&\left|\frac{1+k^2+2k e^{-2i\varphi}}{1-k^2}\right|^{2}, \hspace{0.3cm} 0\leq k<1,
\end{eqnarray}
which means that the quantum conversion rate from the optical to microwave field equals to that from the  microwave to optical field. Once again we observe the reversible quantum conversion between the optical and microwave fields.

In particular, when $\varphi=0$  from Eq. (30) we have
\begin{eqnarray}
\eta_{ow}(\pi/4, \varphi=0)&=&\eta_{wo}(\pi/4, \varphi=0) \nonumber\\
&=&\left(\frac{1+k}{1-k}\right)^{2}, \hspace{0.5cm} 0\leq k<1,
\end{eqnarray}
which indicates that the optical-microwave conversion rate can be sensitively enhanced with the increase of the coupling ratio $k$.  Hence we can observe the reversible entanglement-enhanced quantum conversion between the optical and microwave fields  in this situation.

In order to assess the effect of the initial entanglement on the optical-microwave quantum conversion, we introduce a characteristic parameter, the entanglement-affecting factor (EAF), which is defined by the following expression
\begin{equation}
R(\varphi)=\frac{\eta_{ow}(0, \varphi)}{\eta_{ow}(\pi/4, \varphi)}=\frac{\eta_{wo}(0, \varphi)}{\eta_{wo}(\pi/4, \varphi)},
\end{equation}
which is the ratio of the optical-microwave conversion rate  without the initial entanglement with respect to that with maximal initial entanglement. $R(\varphi)< 1$ implies that the initial optical-microwave entanglement can enhance the optical-microwave conversion rate while $R(\varphi)> 1$ means that that the initial entanglement can suppress the optical-microwave conversion. When  $R(\varphi)= 1$, the optical-microwave conversion rate with the initial entanglement equals to that without the initial entanglement, so that the initial entanglement does not affect the optical-microwave conversion rate in this situation.

The optical-microwave conversion rate with the maximal initial entanglement $\eta_{wo}(\theta=\pi/4, \varphi)$ is given by Eq.(30) while the optical-microwave conversion rate without initial entanglement can be obtain from Eq. (28) by setting $\theta=0$ or $\pi/2$ with the following expression
\begin{equation}
\eta_{ow}(0, \varphi)=\eta_{wo}(0, \varphi)=\left(\frac{2k}{1-k^2}\right)^2,
\end{equation}
which means that the optical-microwave conversion rate without initial entanglement is independent of the phase of the initial-state parameter $z$.

When  the initial-state parameter $z$ is a real number with $\varphi=0$, from Eqs. (31) and (33) we can find that the EAF is given by
\begin{equation}
R(0)=\frac{\eta_{ow}(0, 0)}{\eta_{ow}(\pi/4, 0)}= \left(\frac{\sqrt{2k}}{1+k}\right)^4<1,
\end{equation}
which indicates that the initial-state entanglement enhances the quantum conversion rate in the whole regime of $0\leq k<1$.

When the initial-state parameter $z$ is a pure imaginary number with $\varphi=\pi/2$,  from Eq. (30) we obtain
\begin{equation}
\eta_{ow}(\pi/4, \pi/2)=\left(\frac{1-k}{1+k}\right)^2.
\end{equation}
Then we can get the following expression of the EAF
\begin{equation}
R(\pi/2)=\frac{\eta_{ow}(0, \pi/2)}{\eta_{ow}(\pi/4, \pi/2)}= \left(\frac{\sqrt{2k}}{1-k}\right)^4.
\end{equation}

\begin{figure}[htp]
\centering
\includegraphics[width=8.0cm,height=5.5cm]{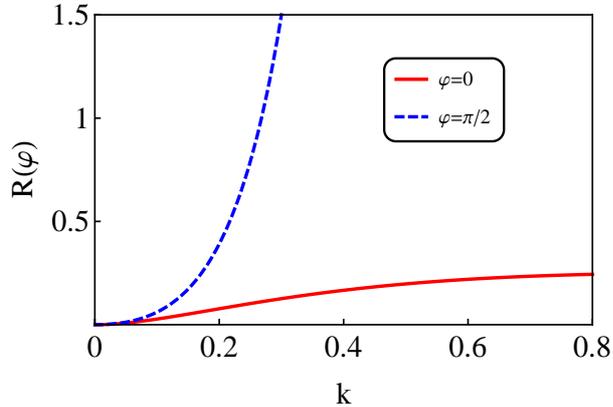}
\caption{(Color online)  The entanglement-affecting factor $R(\varphi)$ with respect to the coupling ratio $k$. The solid and dashed lines correspond to the two cases with and without the initial optical-microwave entanglement, respectively.} \label{fig4}
\end{figure}

In Fig. 4 we have plotted  the EAF  $R(\varphi)$  with respect to the coupling ratio $k$. The solid and dashed lines correspond to the two cases with and without the initial entanglement ($\varphi=0$, $\pi/2$), respectively.   From Fig. 4 we can find that the initial entanglement in the case of  $\varphi=\pi/2$ can exhibit different influence on  the optical-microwave conversion in different regimes of the coupling ratio $k$. From Eq. (36) it is easy to see that there exist a critic point of the coupling ratio $k_c$. The initial entanglement can enhance the optical-microwave conversion in the regime of $k<k_c$ with $R(\varphi)<1$ while the initial entanglement can suppress  the optical-microwave conversion in the regime of $k>k_c$ with $R(\varphi)>1$. However, the initial entanglement does not affect the optical-microwave conversion rate at the critic point $k=k_c$. Making use of Eq. (36) we can find the critic point value  $k_c=2-\sqrt{3}$.

Above analyses indicate that the quantum conversion between the optical wave and microwave can be efficiently manipulated by controlling the initial-state parameters. In particular, we find that      the entanglement-assisted quantum conversion is a kind of the phase-sensitive quantum conversion under certain condition.

\section{Concluding remarks}
We have proposed the DDM  scheme to realize reversible and highly efficient quantum conversion between the microwave and optical fields in terms of the EOM quantum conversion model. We obtain an analytical solution of the EOM model by means of the Senm-Mandal approch. It is demonstrated that two optomechnical DDMs appear at some specific moments during the dynamical evolution of the EOM model. It is found that the DDMs can induce two kinds of  bidirectional and highly efficient quantum conversion. The first one is the microwave-optical CQC which happens at the condition of vanishing of the initial-state mean value of one of the microwave and optical fields. The CQC is reversible since the bidirectional quantum conversion has the same conversion rate in this case. In some sense the DDM-assisted reversible quantum conversion is universal because the  bidirectional conversion rate is independent of the initial state of the converted field, it only depends on the coupling ratio between the microwave and optical fields and the mechanical resonator. The second one is the EAQC which occurs in the presence of the initial-state entanglement  between the microwave and optical fields. We have demonstrated  that  the EAQC  between microwave and optical fields can be manipulated by varying the initial-state entanglement and the coupling ratio of related interaction strengths. Especially, it has been indicated the  microwave-optical EAQC  is the phase sensitive conversion. It is found that it is possible to realize the entanglement-enhanced or entanglement-suppressed quantum conversion through controlling the phase of the initial-state parameter.
It should be mentioned that the DDM scheme is robust against the mechanical noise since the DDMs can protect the system from mechanical dissipation through the two DDMs decoupling with the mechanical mode of the system under consideration.
It should be noted that the reversible and highly efficient quantum conversion is a DDM effect essentially.
It could be expected that the DDMs may lead to more novel quantum phenomena which deserve to be further explored.
The DDM-assisted quantum conversion scheme proposed in the present paper provides a versatile route to manipulate the microwave-optical quantum conversion with the DDMs.
The ability to coherently convert information between microwave and optical fields opens new possibilities for quantum information, in particular,  quantum-coherent connection between microwave and optical photons mediated by the mechanical resonator.

\begin{acknowledgements}
{This work is supported by the National Natural
Science Foundation of China under Grants No. 11775075 and No. 11935006, the STI Program of Hunan Province under Grant No. 2020RC4047.}
\end{acknowledgements}



\end{CJK} 

\end{document}